\documentclass[%
 reprint,
preprintnumbers,
longbibliography,
 amsmath,amssymb,
 aps,
 prc,
 superscriptaddress,
]{revtex4-2}

\usepackage{xcolor}
\usepackage{graphicx} 
\usepackage{subfigure}
\usepackage{dcolumn} 
\usepackage{bm}
\usepackage{float}

\newcommand{\mean}[1]{\langle #1 \rangle}

\newcommand{\psirp}{\Psi^{\rm  RP}}

\newcommand{\psipp}{\Psi^{\rm  PP}}
\newcommand{\psisp}{\Psi^{\rm  SP}}

\usepackage{hyperref}
\usepackage[mathlines]{lineno}
\usepackage{color}
\definecolor{orange}{cmyk}{0.,0.353,1.,0.}    

\begin{document}

\title{Flow-plane decorrelations in heavy-ion collisions with multiple-plane cumulants}

\author{Zhiwan Xu}\email{zhiwanxu@physics.ucla.edu}
\affiliation{Department of Physics and Astronomy, University of
  California, Los Angeles, California 90095, USA}

\author{Xiatong Wu} \affiliation{Department of Physics and Astronomy,
  University of California, Los Angeles, California 90095, USA}

\author{Caleb Sword} \affiliation{Department of Physics and Astronomy,
  Wayne State University, 666 W. Hancock, Detroit, Michigan 48201 }

\author{Gang Wang}\email{gwang@physics.ucla.edu}
\affiliation{Department of Physics and Astronomy, University of
  California, Los Angeles, California 90095, USA} 

\author{Sergei A. Voloshin} \affiliation{Department of Physics and
  Astronomy, Wayne State University, 666 W. Hancock, Detroit, Michigan
  48201 } 

\author{Huan Zhong Huang} \affiliation{Department of Physics and
  Astronomy, University of California, Los Angeles, California 90095,
  USA} \affiliation{Key Laboratory of Nuclear Physics and Ion-beam
  Application (MOE), and Institute of Modern Physics, Fudan
  University, Shanghai-200433, People’s Republic of China}



\begin{abstract}

The azimuthal correlations between local flow planes at different
(pseudo)rapidities ($\eta$) may reveal important details of the
initial nuclear matter density distributions in heavy-ion collisions.
Extensive experimental measurements of a
factorization ratio ($r_2$) and its derivative ($F_2$) have shown
evidence of the longitudinal flow-plane decorrelation. However,
nonflow effects also affect this observable and prevent a quantitative understanding of the phenomenon. In this paper, to distinguish decorrelation and
nonflow effects, we propose a new cumulant observable, $T_2$, which
largely suppresses nonflow.  The technique sensitivity to different initial-state
scenarios and nonflow effects are tested with a simple Monte Carlo
model, and in the end, the method is applied to events simulated by a multiphase
transport model (AMPT) for Au+Au collisions at $\sqrt{s_{\rm NN}} =200$~GeV.
{\color{black} We also emphasize that a distinct decorrelation signal requires not only the right sign of an observable, but also its proper dependence on the $\eta$-window of the reference flow plane, to be consistent with the pertinent decorrelation picture.}
 
\begin{description}
\item[keywords]
flow decorrelation, nonflow
\end{description}
\end{abstract}

\maketitle
\section{Introduction}
\label{sec:intro}

\begin{figure}
\includegraphics[width=0.48\textwidth]{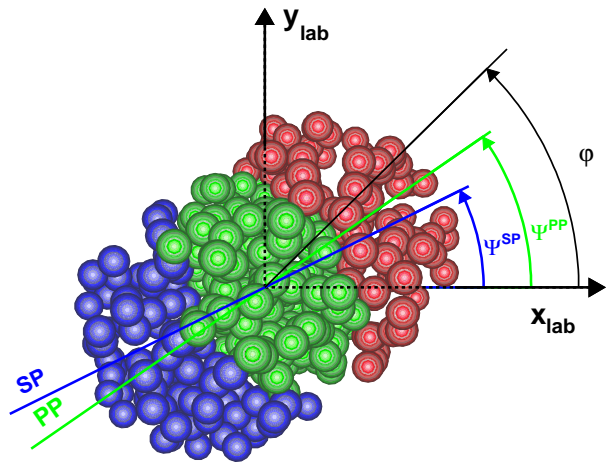}
\caption{\label{fig:col_setup} (Color online) Schematic view of a two-nucleus collision in the transverse plane. The left nucleus is emerging
  from and the right one going into the page. Particles are produced
  in the overlap region (green-colored are the participant
  nucleons). The azimuthal angles of the spectator plane ($\psisp$),
  the participant plane ($\psipp$) and one of the produced particles ($\varphi$)
  are depicted here.  }
\end{figure}

Experiments on high-energy heavy-ion collisions, such as those at the
BNL Relativistic Heavy Ion Collider (RHIC) and the CERN Large Hadron Collider
(LHC), aim to create a quark gluon plasma (QGP) and to study the
properties of this deconfined nuclear medium.  Most heavy-ion collisions
are not head-on, and traditionally, the nucleons experiencing at least
one collision are considered as participants, and the remaining are
labeled as spectators (see Fig.~\ref{fig:col_setup}).  While spectators
fly away, the system created by the participant
interaction presumably undergoes a hydrodynamic expansion. The initial
geometry of the system is determined by the participant distribution,
with event-by-event fluctuations.  The pressure gradients of
the medium convert the spatial anisotropies of the initial matter
distribution into the momentum anisotropies of the final-state
particles.  Consequently, the azimuthal distributions of emitted
particles can be analyzed with a Fourier
expansion~\cite{Voloshin_1996,Methods}
\begin{equation}
\frac{dN}{d\varphi} \propto 1 + \sum_{n=1}^{\infty} 2v_n \cos[
  n(\varphi -\psirp)],
\label{equ:Fourier_expansion}   
\end{equation}
where $\varphi$ denotes the azimuthal angle of a particle and
$\psirp$ is the reaction plane azimuth (defined by the impact
parameter vector). The Fourier coefficients,
\begin{equation}
v_n = \langle \cos [n(\varphi -\psirp)] \rangle \,,  
\label{eq:expansionRP}
\end{equation}
are referred to as anisotropic flow of the $n^{\rm th}$ harmonic.  By
convention, $v_1$, $v_2$ and $v_3$ are called ``directed flow",
``elliptic flow", and ``triangular flow", respectively.  They reflect
the hydrodynamic response of the system to the initial geometry (and
its fluctuations) of the participant zone~\cite{HYDRO_review}.

In reality, the reaction plane is unknown, and more importantly,  
the initial-state fluctuations drive the anisotropic flow  along the planes that differ from the reaction plane, the so-called flow 
symmetry planes or participant planes ($\Psi_n^{\rm PP}$). 
Then the particle azimuthal distributions can be rewritten as
\begin{equation}
\frac{dN}{d\varphi} \propto 1 + \sum_{n=1}^{\infty} 2v_n \cos[
  n(\varphi -\Psi_n^{\rm PP})].
\label{eq:expansionPP}   
\end{equation}
The meaning of the flow coefficients $v_n$ changes from those in
Eq.~\ref{eq:expansionRP}, but for simplicity of notations, the same symbols will be used, since in later discussions we will not determine the flow coefficients with the reaction plane. Anisotropic flow measurements relative to the participant plane are
straightforward, as the flow itself can be used to estimate the corresponding flow plane. However, using the participant/flow plane also
has its drawback -- these planes become dependent on the kinematic region (rapidity and transverse momentum) of particles involved. This dependence is
relatively weak, which still justifies the flow formalism of
Eq.~\ref{eq:expansionPP}, but it needs to be taken into account to interpret high-precision flow measurements in modern experiments, {\color{black} especially the flow-plane decorrelation analyses to be discussed.}

For clarity, we collect the definitions of different planes used in
this paper below:
\begin{itemize}
\item 
Reaction plane (RP) is the plane spanned by the beam direction and the
impact parameter vector. This plane is unique for every collision.
\item
Participant plane (PP) is defined by the initial density
distribution. Subtle differences may exist, depending on, e.g., whether
entropy or energy density is used as a weight, but these potentially
small differences are not discussed in this paper. We assume that the properly constructed PPs  define the development of anisotropic
flow.
\item
Flow symmetry plane or flow plane (FP) determines the orientation of the corresponding harmonic anisotropic
flow.  It is assumed that FP coincides with the PP of
the same harmonic (linear flow mode) or a proper combination of the
lower harmonic PPs (nonlinear flow mode). With the nonlinear flow
modes neglected, FP and PP are often used interchangeably.
\item 
Event plane (EP) estimates the FP by analyzing the particle azimuthal distribution in a particular kinematic region. Owing to the finite number of particles involved in 
such an estimate, EP is  subject to statistical fluctuations. The 
measurements obtained with EP have to be corrected for the
event plane resolution~\cite{Methods}, characterized by $\langle\cos[n(\Psi_n^{\rm EP}-\Psi_n^{\rm FP})]\rangle$.
$\Psi^{\rm EP}$ is the azimuthal angle of the reconstructed $n^{\rm th}$-harmonic flow vector, ${\bf Q}_n =(\sum_i^N
w_i\cos(n\varphi_i),\sum_i^N w_i\sin(n\varphi_i))$, where $w_i$ is the
weight for each particle.  For simplicity, we use unity weights in the
event plane calculation.
\item
Spectator planes (SP) is determined by a sideward deflection of spectator nucleons, and is regarded as a better proxy for RP than FPs (determined by participants).
\end{itemize}

\begin{figure}
\includegraphics[width=0.48\textwidth]{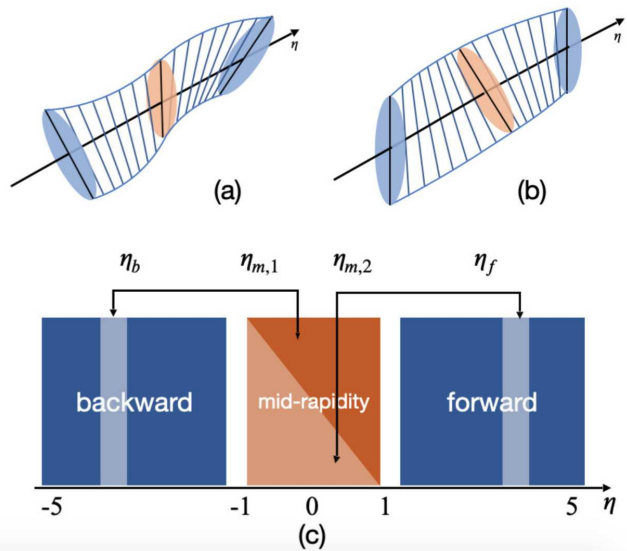}%
\caption{\label{fig:de_setup} (Color online) Schematic view of (a) the ``torque" or S-shaped , and (b) the bow or C-shaped  decorrelation patterns in the longitudinal distribution of the flow
plane angle. The ellipses indicate the transverse momentum distributions of the final-state particles. 
Panel (c) delimits the
kinematic regions for the particles at mid-, forward, and backward   pseudorapidities.}
\end{figure}

The objective of this paper is the flow-plane
decorrelation in the (pseudo)rapidity ($\eta$) direction. Decorrelation here means the deviation of a local flow plane from the value at the center-of-mass rapidity ($y_{\rm CM}$), 
$\Delta\Psi(\eta)=\Psi(\eta)-\Psi(y_{\rm CM})$. For simplicity, $y_{\rm CM}$ is set to zero for the symmetric collisions under study.
In practice, we measure the relative tilt angles between the flow planes at backward, mid-, and forward rapidities. For concreteness we
focus on the $2^{\rm nd}$-harmonic flow. From event to event, two possible  patterns arise from flow
fluctuations: (a) when the flow plane angles
at forward and backward pseudorapidities ($\Psi_{\rm f}$ and
$\Psi_{\rm b}$) fall on the opposite sides of the flow plane at
midrapidities ($\Psi_{\rm m}$) -- the ``torque'' scenario,
or S-shaped decorrelations, and (b) when $\Psi_{\rm f}$ and
$\Psi_{\rm b}$ fall on the same side relative to $\Psi_{\rm m}$ -- the ``bow" scenario, or C-shaped decorrelations. These two cases are exemplified in panels (a) and
(b) of Fig.~\ref{fig:de_setup}, respectively.

The magnitude and pattern of the flow-plane decorrelation  is extremely important, not only for the flow measurements (to be discussed in Sec.~\ref{Sec:v2}), 
but also for understanding of the initial condition in the longitudinal direction. Flow-plane decorrelations can be caused by the torque
effect~\cite{torque}, and more generally eccentricity
decorrelations~\cite{EccenDecorr}. The mechanisms leading to the decorrelations also include hydrodynamic fluctuations in the QGP
fluid~\cite{HydroFluc} and glasma dynamics~\cite{Glasma}. There also exists a
phenomenological dynamical model of the initial states~\cite{Dynamic} that predicts the torque. 
We cannot exclude the possibility that
the mechanisms causing the S-shaped decorrelations coexist with those originating the C-shaped ones in heavy-ion collisions.
Therefore, experimental observables are only expected to measure 
the average effect and reveal the dominant decorrelation pattern.

A widely used measure of the longitudinal flow-plane decorrelation
was introduced by the CMS Collaboration~\cite{rn}:
\begin{eqnarray}
r_n(\eta) 
&=& \frac{\langle \cos[n(\varphi_{-\eta} - \Psi_{\rm f})]
  \rangle}
{\langle \cos[n(\varphi_{\eta} - \Psi_{\rm
      f})]\rangle} \label{eq:3} 
\\
&=& \frac{\langle v_n(-\eta) \cos[n(\Psi_{-\eta}-\Psi_{\rm f})]\rangle}
{\langle v_n(\eta) \cos[n(\Psi_{\eta}-\Psi_{\rm f})]\rangle}, \label{eq:4}
\label{equ:r_n}
\end{eqnarray}
with $\eta >0$. $\Psi_{\rm f}$ can be replaced with 
$\Psi_{\rm b}$, if $\eta$ and $-\eta$ are simultaneously swapped in the definition. In a
symmetric collision without any decorrelation, one would expect that $v(\eta)=v(-\eta)$ and the
ratio to be unity. But if a torque pattern is present,
with $\Psi_{\rm f}$ serving as a reference point, the decorrelation effect would be stronger for the negative-rapidity region,
and the ratio would go below unity.  Experimental data show that the factorization ratio
$r_2$ indeed decreases with increasing $\eta$, and the deviation from
unity is typically a few percent per unit  pseudorapidity at both
the LHC~\cite{rn,ATLAS,ATLAS2} and the RHIC~\cite{Maowu}. Since both the
flow-plane decorrelation and the flow-magnitude decorrelation can
cause such a dependence of $r_2$ on $\eta$, efforts have been made to separate the two
contributions~\cite{ATLAS,Bozek,Legendre}. However, before relating
the observed $r_2(\eta)$ dependence to the flow-plane decorrelation,
one has to examine an important physics background, the nonflow.

The nonflow effects are the correlations unrelated to
the flow plane orientation or the initial geometry. Some nonflow
effects are short-range in pseudorapidity, such as Coulomb and
Bose-Einstein correlations (a few tenths of the unit of rapidity),
resonance decays, and intra-jet correlations (about 1 unit rapidity), whereas back-to-back jets could contribute to the
long-range correlations spanning over several units of rapidity.
Therefore, even with a sizable $\eta$ gap between an event plane and
the particles of interest, one can not completely eliminate nonflow
contributions to $v_2$ or $r_2$ measurements.  Nonetheless, the
numerator of $r_2$ does involve a larger $\eta$ gap and hence a
smaller nonflow contribution than its denominator, leading to a ratio
smaller than unity, similar to that caused by the S-shaped flow-plane
decorrelation.
On the other hand, the C-shaped decorrelation may also be faked in experimental observables by nonflow effects such as back-to-back jet correlations.

In general, the decorrelation observables also depend on the $\eta$-window of the reference flow plane. In the following sections, we  examine this dependence for various observables in the presence of nonflow and different types of flow-plane decorrelations.
In Sec II, we use simple Monte Carlo simulations to demonstrate the impact of flow-plane decorrelations and nonflow on the $v_2$ measurements.
Section III discusses the $r_2$ (and the closely related $F_2$ observable) analyses, and  illustrates the possibilities of the
$\eta$-differential measurements for a better interpretation of the
results.  We  show that nonflow effects tend to cause overestimation of $F_2$, and to distort the $|\eta_{\rm f(b)}|$-dependence
originally created by the flow-plane decorrelations. 
In Sec IV, we introduce a
new four-plane observable, $T_2$, which is essentially free from 
the nonflow contribution and has very distinct expectations for different
decorrelation patterns.  In Sec V, the developed techniques are applied to
the Au+Au events generated by a multiphase transport (AMPT) model~\cite{ampt_1}. 
Finally in Sec VI, we summarize the findings and
discuss the application of the new method to experimental data.

\section{\texorpdfstring{$v_2$}{Lg} Monte-Carlo simulations}
\label{Sec:v2}

In elliptic flow measurements, it is a common practice to introduce a
sizeable $\eta$ gap between the event plane ($\Psi_{\rm f(b)}$) and
the particles of interest ($\varphi_{\rm m}$) to suppress nonflow:
\begin{equation}
v_2\{\Psi_{\rm f(b)}\} = \frac{\langle \cos[2(\varphi_{\rm m} 
- \Psi_{\rm  f(b)})]\rangle}
{\sqrt{\langle \cos[2(\Psi_{\rm f} - \Psi_{\rm  b})]\rangle}},
\label{eq:v2_EP}
\end{equation}
where the denominator is the event plane resolution for $\Psi_{\rm
  f(b)}$.
  Although both the numerator and the denominator are
contaminated by nonflow, the latter is less affected because of a larger $\eta$ gap between $\Psi_{\rm f}$ and $\Psi_{\rm b}$. 
Therefore the effect of nonlow on $v_2\{\Psi_{\rm f(b)}\}$ decreases 
with increasing the $\eta$ gap or $|\eta_{\rm f(b)}|$.

\begin{figure}[b]
\includegraphics[width=0.48\textwidth]{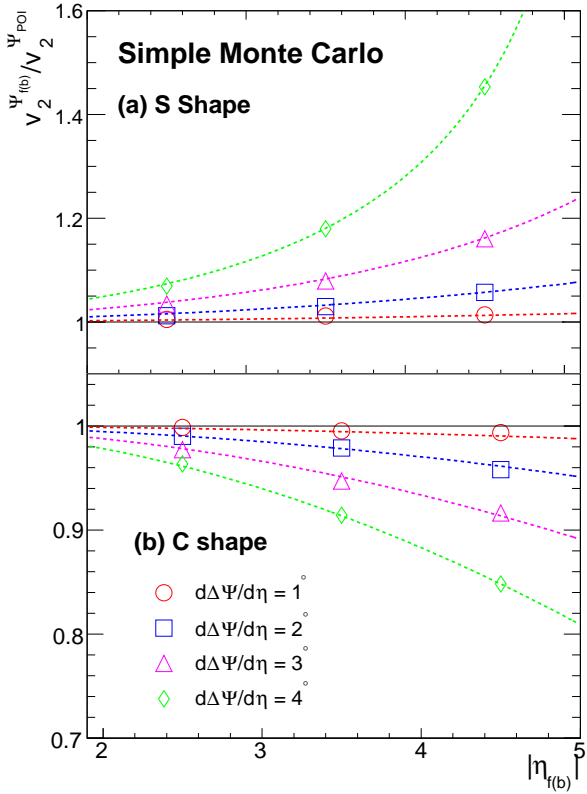}
\caption{\label{fig:v2_MC} (Color online) Simple Monte Carlo
  calculations of the ratio, $v_2\{\Psi_{\rm f(b)}\}/v_{2}\{\Psi_{\rm POI}\}$, without nonflow
  for the S-shaped (a) and the C-shaped (b) flow-plane decorrelations as a
  function of $|\eta_{\rm f(b)}|$ and $\frac{d\Delta\Psi}{d\eta}$.
  The curves follow Eqs.~(\ref{eq:Sshape}) and (\ref{eq:Cshape}).}
\end{figure}

When the S-shaped flow-plane decorrelation is present, 
both the numerator and the
denominator in Eq.~\ref{eq:v2_EP} are reduced by the finite tilt
angles due to the $\eta$ gaps, and the denominator is more
influenced because it involves an  $\eta$ gap larger than that of
the numerator. Therefore, the S-shaped decorrelation tends to drive
$v_2\{\Psi_{\rm f(b)}\}$ above $v_2\{\Psi_{\rm POI}\}$, where $\Psi_{\rm POI}$ is the event plane determined in the same $\eta$ region as the particles of interest (POI).

To test these speculations,
we perform a simple Monte Carlo simulation, where
the tilt angle, $\Delta\Psi(\eta)$, is a linear function of $\eta$, with a slope of $\frac{d\Delta\Psi}{d\eta} =
1^{\circ}$, $2^{\circ}$, $3^{\circ}$ or $4^{\circ}$:
\begin{equation}
\Delta\Psi(\eta) \equiv  \Psi(\eta) - \Psi(0) = \frac{d\Delta\Psi}{d\eta} \eta.
\label{eq:sshape}
\end{equation}

In our simulation (see Appendix A for details), 
each event has  1000 particles, distributed  uniformly in the $\eta$ range of 
(-5, 5). 
The particle density $\frac{dN}{d\eta}|_{\eta=0}=100$ corresponds roughly to the 30--50\% centrality range in Au+Au collisions at $\sqrt{s_{\rm NN}} = 200$ GeV, or 50--70\% Pb+Pb at $\sqrt{s_{\rm NN}} = 2.76$ TeV.
The azimuthal
angle of each particle has been assigned randomly according to the distribution of Eq.~\ref{eq:expansionPP},
and $\Psi^{\rm PP}$ is replaced with $\Psi(\eta)$.
The POIs are selected
within $|\eta|<1$, and the $|\eta_{\rm f(b)}|$ ranges used for calculations of $\Psi_{\rm f}$ and $\Psi_{\rm b}$ are taken from three
bins: (2, 3), (3, 4) and (4, 5). For simplicity, the input $v_2$ is independent of $\eta$ or transverse momentum ($p_T$), and the simulation does not include nonflow at this step. We implement a $15\%$ flow fluctuation in all the following simple simulations, and observe no difference from results with zero fluctuation.
Figure~\ref{fig:v2_MC}(a)
shows that 
the $v_2\{\Psi_{\rm f(b)}\}/v_{2}\{\Psi_{\rm POI}\}$ ratio is above unity, and increases with both $|\eta_{\rm f(b)}|$ and $\frac{d\Delta\Psi}{d\eta}$, as expected for the S-shaped  decorrelations. 
In the case of the C-shaped
decorrelations, similar to Eq.~\ref{eq:sshape}, we assume
\begin{equation}
\Delta\Psi(\eta) = \frac{d\Delta\Psi}{d\eta} |\eta|,
\label{eq:cshape}
\end{equation}
and still vary $\frac{d\Delta\Psi}{d\eta}$  from $1^{\circ}$ to
$4^{\circ}$.  Now that the tilt angle is symmetric around $\eta=0$,
the denominator in Eq.~\ref{eq:v2_EP} is unchanged, and
$v_2\{\Psi_{\rm f(b)}\}$ tends to under-estimate $v_2\{ \Psi_{\rm POI}\}$. Indeed, Fig.~\ref{fig:v2_MC}(b) shows that the 
$v_2\{\Psi_{\rm f(b)}\}/v_2\{ \Psi_{\rm POI}\}$ ratio goes below unity, and decreases with
increasing $|\eta_{\rm f(b)}|$ and $\frac{d\Delta\Psi}{d\eta}$ for the C-shaped case. The ratios for the S-shaped and C-shaped scenarios can
be well described, respectively, by
\begin{eqnarray}
\frac{v_2\{\Psi_{\rm f(b)}\}}{v_{2}\{\Psi_{\rm POI}\}} &=&
\frac{\cos(2\frac{d\Delta\Psi}{d\eta} |\eta_{\rm
    f(b)}|)}{\sqrt{\cos(4\frac{d\Delta\Psi}{d\eta} |\eta_{\rm
      f(b)}|)}} {\rm
  ~(S~shape)}, \label{eq:Sshape}\\ \frac{v_2\{\Psi_{\rm
    f(b)}\}}{v_{2}\{\Psi_{\rm POI}\}} &=& \cos[2\frac{d\Delta\Psi}{d\eta}(
  |\eta_{\rm f(b)}|-0.5)] {\rm~(C~shape)} \label{eq:Cshape}.
\end{eqnarray}
Therefore, to manifest a clear decorrelation signal, the $v_2$ ratio should go above unity with a rising trend vs $|\eta_{\rm f(b)}|$ for a torque (S-shaped) case, or below unity with a falling trend for a bow (C-shaped) case.

\begin{figure}[tb]
\includegraphics[width=0.48\textwidth]{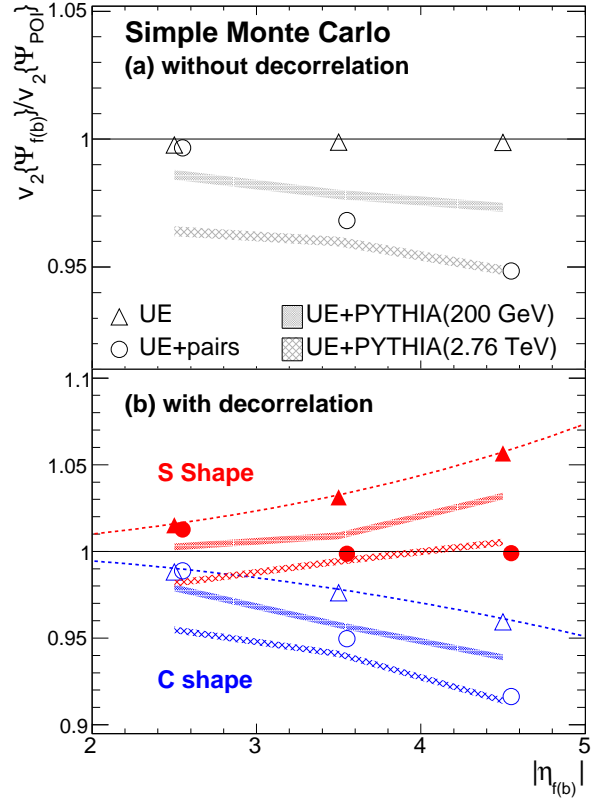}
\caption{\label{fig:v2_nonflow} (Color online) Simple Monte Carlo calculations of the $v_2\{\Psi_{\rm f(b)}\}/v_{2}\{\Psi_{\rm POI}\}$ ratio as a function of $|\eta_{\rm f(b)}|$ {\color{black} for the underlying event (UE) without nonflow as well as for different scenarios of nonflow}.
$\frac{d\Delta\Psi}{d\eta}$ is set to $2^\circ$  for the S-shaped and the C-shaped flow-plane decorrelations.
The curves follow  Eqs.~\ref{eq:Sshape} and \ref{eq:Cshape} without nonflow. {\color{black}The shaded bands represent the two cases of PYTHIA embedding.}}
\end{figure} 

We further implement  nonflow correlations in the simulation, by adding to the ``underlying" event (UE) pairs of particles with the same azimuthal angle.
This nonflow simulation is similar to that used in Sec. IV(D) of Ref~\cite{Aihong}.
Note that since we deal with the $2^{\rm nd}$-harmonic flow, the near-side pairs with $\Delta\varphi\sim 0$ are equivalent to the away-side pairs with $\Delta\varphi\sim \pi$.
One of the paired particles follows the uniform $\eta$ distribution within (-5, 5), and the other particle is separated with an $\eta$ gap that obeys a Gaussian distribution with a width of 2 units of  pseudorapidity.
Four hundred paired particles (200 pairs) have been added to each underlying event.
The simulated $v_2\{\Psi_{\rm f(b)}\}/v_{2}\{\Psi_{\rm POI}\}$ ratio is shown in Fig.~\ref{fig:v2_nonflow} as a function of $|\eta_{\rm f(b)}|$ with different amounts of nonflow.
Panel (a) displays the scenario without any decorrelation, where nonflow fakes a falling trend.
Panel (b) gives an example with $\frac{d\Delta\Psi}{d\eta} = 2^{\circ}$ for the S-shaped and the C-shaped flow-plane decorrelations.
In both scenarios, nonflow pulls down the original trends,
and in the case of the S-shaped decorrelation, the initially rising trend could even be reversed into a falling one.
Therefore, the $v_2\{\Psi_{\rm f(b)}\}/v_{2}\{\Psi_{\rm POI}\}$ ratio alone lacks discernment of different decorrelation scenarios. 

Besides the simplified implementation of nonflow, which could exaggerate the effect, we also take a more realistic approach by embedding a few PYTHIA~\cite{PYTHIA} events from p+p collisions at $\sqrt{s}= 200$ GeV or $2.76$ TeV, such that PYTHIA particles replace the 400 paired particles (200 pairs). PYTHIA is an event generator that comprises a coherent set of physics mechanisms for the evolution from a few-body hard
scattering process to a complex multihadronic final state. The corresponding results are shown with shaded bands in Fig.~\ref{fig:v2_nonflow}. The $v_2$ ratios thus obtained qualitatively show pull-down effects similar to the simplified nonflow case, with a stronger magnitude at the higher collision energy. In this study the embedding of PYTHIA particles into underlying events is done mostly to illustrate the effects of nonflow, but with parameters tuned to a specific centrality interval, it can also provide a quantitative estimate of  nonflow contributions in the data analyses. More discussions on the simple Monte Carlo simulations and  the nonflow effects can be found in Appendix~\ref{Append:Nonflow}.

\section{\texorpdfstring{$r_2$}{Lg} and \texorpdfstring{$F_2$}{Lg}}

We define $r_2(\eta)$ based on Eq.~\ref{eq:3} by setting $n = 2$. As
suggested by Eq.~\ref{eq:4}, the deviation of $r_2$ from unity may
originate from the decorrelations both in the flow-plane angles and in the $v_2$ magnitudes.  Thus, we also examine the
modified observable, $r_2^{\Psi}$~\cite{Bozek}, which is supposedly
sensitive only to the flow-plane angles:
\begin{equation}
r_2^\Psi(\eta) = \frac{\langle \cos[2(\Psi_{-\eta}-\Psi_{\rm f})]\rangle} {\langle \cos[2(\Psi_{\eta}-\Psi_{\rm
f})]\rangle}.
\end{equation}
Experimentally as well as in model studies below, the dependence of $r_2^{(\Psi)}$ on
$\eta$ is almost linear, and the $F_2^{(\Psi)}$ slope is used to quantify
the effect~\cite{rn}:
\begin{eqnarray}
    r_2^{(\Psi)} = 1 - 2 F_2^{(\Psi)} \eta. 
\end{eqnarray}
%

\begin{figure}[tb]
\includegraphics[width=0.48\textwidth]{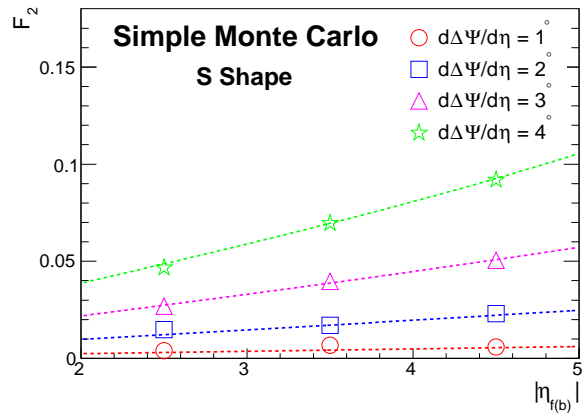}
\caption{\label{fig:F2_MC} (Color online) $F_2$ as a function of
  $|\eta_{\rm f(b)}|$ and $\frac{d\Delta\Psi}{d\eta}$, from a
  simple Monte Carlo simulation for the S-shaped flow-plane
  decorrelation without nonflow. The curves reflect the analytical relation in Eq.~\ref{eq:F2_eta}.  }
\end{figure}

We perform the linear-$\Delta\Psi(\eta)$ Monte Carlo simulation without nonflow to inspect the qualitative
expectation of $F_2$ in the presence of the S-shaped flow-plane
decorrelation.  Note that in the C-shaped case, $F_2$ is zero by
construction. In our simple simulations, 
$F_2$ and $F_2^\Psi$ are always identical, so only the $F_2$ results are presented. Figure~\ref{fig:F2_MC} depicts $F_2$
as a function of $|\eta_{\rm f(b)}|$ and $\frac{d\Delta\Psi}{d\eta}$.  
$F_2$ increases with $\frac{d\Delta\Psi}{d\eta}$, since a larger tilt angle means a stronger torque.  At
first glance, it seems to be counter-intuitive that $F_2$ depends on
the $\eta$ location of the reference event plane, but the simulation actually
reveals a simple mathematical relation:
\begin{equation}
1 - 2F_2 \eta =  \frac{\cos[2(\eta+|\eta_{\rm f(b)}|)\frac{d\Delta\Psi}{d\eta}]} 
{ \cos[2(\eta-|\eta_{\rm f(b)}|)\frac{d\Delta\Psi}{d\eta}]}
, \label{eq:F2_eta}
\end{equation}
which yields $F_2\approx 4 (d\Delta\Psi/d\eta)^2|\eta_{\rm f(b)}|$.  Although in
real collisions, the dependence of $\Delta\Psi$ on $\eta$ may not be linear, 
we have verified with various monotonic function forms  that
the larger the $\eta$ gap between POIs and the reference
event plane is, the larger $F_2$ is.  Thus an experimental observation of {\color{black} positive $F_2$ values with an}  increasing
$F_2(|\eta_{\rm f(b)}|)$ trend  {\color{black} may reveal a distinct domination} of the S-shaped flow-plane decorrelations.

\begin{figure}[tb]
\includegraphics[width=0.48\textwidth]{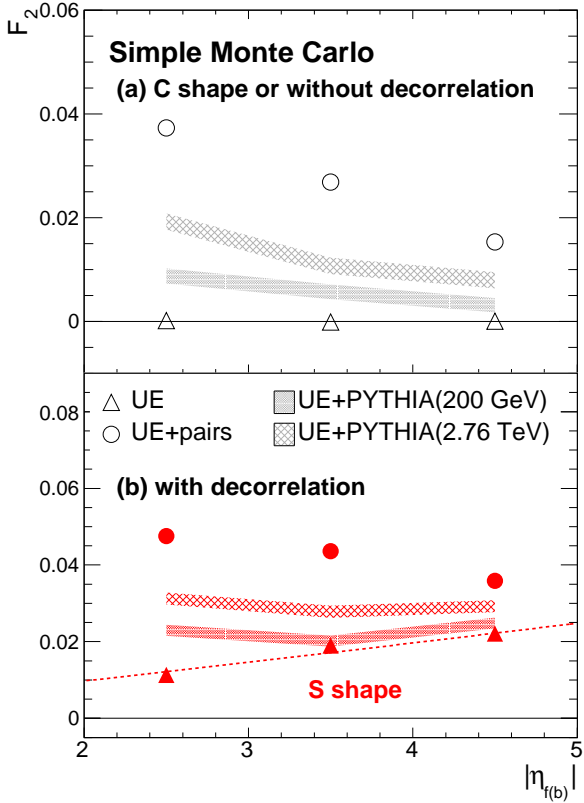}
\caption{\label{fig:F2_nonflow} (Color online) Simple Monte Carlo calculations of $F_2(|\eta_{\rm f(b)}|)$ {\color{black} for the underlying event (UE) without nonflow as well as for different scenarios of nonflow}.
$\frac{d\Delta\Psi}{d\eta}$ is set to $2^\circ$  for the S-shaped and the C-shaped flow-plane decorrelations.
The curve follows the ideal relation  in Eq.~\ref{eq:F2_eta}  without nonflow. {\color{black}The shaded bands represent the two cases of PYTHIA embedding.} }
\end{figure} 

In Fig.~\ref{fig:F2_nonflow}, nonflow contributions have been studied under the same framework as that used for the $v_2\{\Psi_{\rm f(b)}\}/v_{2}\{\Psi_{\rm POI}\}$ ratio. Since the simple simulation results on $F_2$ are the same for the scenarios with no decorrelation and with the C-shaped decorrelation, we use one set of data points to present both of them in panel (a). In these two scenarios, nonflow can fake a finite $F_2$ value, which decreases with increasing $|\eta_{\rm f(b)}|$. Furthermore, panel (b) shows that nonflow not only quantitatively increases the magnitude of $F_2$ for the S-shaped decorrelation, but could also qualitatively change its rising trend into a falling one vs $|\eta_{\rm f(b)}|$. 
{\color{black} The embedding of 400 PYTHIA particles resembles the simplified nonflow implementation with weaker effects, but finite $F_2$ values are still faked when the truth is no decorrelation or the C-shaped decorrelation.
For the S-shaped decorrelation,
the rising trend vs $|\eta_{\rm f(b)}|$ is still distorted, especially at intermediate $|\eta_{\rm f(b)}|$.}
Therefore, $F_2$ cannot unambiguously distinguish  and quantify  
different decorrelation scenarios.
{\color{black}Figure 2(a) of Ref.~\cite{ATLAS2} gives a concrete example of such nonflow effects on the $|\eta_{\rm f(b)}|$ dependence of $F_2$ in Xe+Xe collisions at $\sqrt{s_{\rm NN}} = 5.44$ TeV.}

\section{\texorpdfstring{$T_2$}{Lg} analyses}
\label{Sec:T2}

In view of  possible
significant nonflow contributions in $v_2\{\Psi_{\rm f(b)}\}$ and $F_2$  analyses, we advocate a  new observable to
probe the longitudinal flow-plane decorrelation:
\begin{equation}
T_2 = \frac{\langle \langle \sin 2(\Psi_{\rm f} - \Psi_{{\rm m},1})
\sin 2(\Psi_{\rm b} - \Psi_{{\rm m},2})\rangle \rangle}{ Res(\Psi_{\rm f}) Res(\Psi_{{\rm m},1})  Res(\Psi_{\rm b})  Res(\Psi_{{\rm m},2})},
\label{eq:T2}
\end{equation}
where particles at midrapidities ($|\eta|<1$) are  divided into two sub-events to form $\Psi_{{\rm m},1}$ and $\Psi_{{\rm m},2}$, as
demonstrated in Fig.~\ref{fig:de_setup}(c). The double brackets denote
``cumulant", and operate as follows:
\begin{eqnarray}
& &\langle\langle
\sin(a-b)\sin(c-d)
\rangle\rangle \nonumber\\ &\equiv& \langle
\sin(a-b)\sin(c-d)
\rangle - \frac{1}{2}\langle
\cos(a-c)\rangle\langle\cos(b-d)
\rangle \nonumber\\
& &+ \frac{1}{2}\langle
\cos(a-d)\rangle\langle\cos(b-c)
\rangle,
\end{eqnarray}
the derivation of which is elaborated in Appendix~\ref{Append:cumu}. 
Taking  into account the flow fluctuation contributions in the event plane resolution, we have
\begin{eqnarray}
T_2 &=& \frac{\langle \sin 2(\Psi_{\rm f} - \Psi_{{\rm m},1})
\sin 2(\Psi_{\rm b} - \Psi_{{\rm m},2})\rangle} {\langle \cos 2(\Psi_{\rm f} - \Psi_{{\rm m},1})
\cos 2(\Psi_{\rm b} - \Psi_{{\rm m},2}) \rangle} \nonumber \\
&-&  \frac{\langle \cos 2(\Psi_{\rm f} - \Psi_{\rm b} )\rangle
\langle\cos 2(\Psi_{{\rm m},1} - \Psi_{{\rm m},2})\rangle} {2\langle \cos 2(\Psi_{\rm f} - \Psi_{{\rm m},1}  )\rangle
\langle\cos 2(\Psi_{\rm b} - \Psi_{{\rm m},2}) \rangle} + \frac{1}{2}.
\end{eqnarray}
The generalization of the $T_2$ definition to four independent pseudorapidity ranges is straightforward and is  discussed in Appendix~\ref{Append:general}.

\begin{figure}
\includegraphics[width=0.48\textwidth]{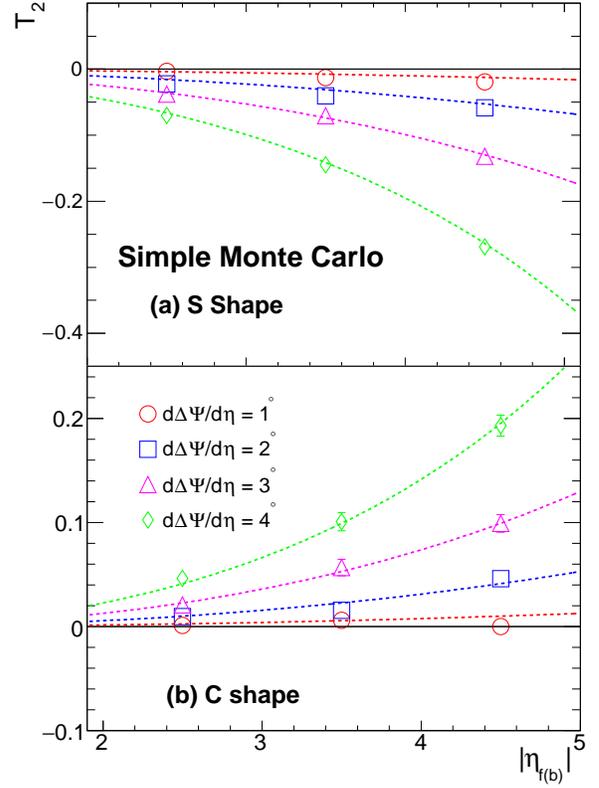}
\caption{\label{fig:T2_MC} (Color online) Simple Monte Carlo
  simulations of $T_2$ as a function of $|\eta_{\rm f(b)}|$ and
  $\frac{d\Delta\Psi}{d\eta}$ for the S-shaped (a) and the C-shaped
  (b) flow-plane decorrelations without nonflow.  The curves express
  Eqs.~\ref{eq:Sshape_T2} and \ref{eq:Cshape_T2}.  }
\end{figure}

\begin{figure}[tb]
\includegraphics[width=0.48\textwidth]{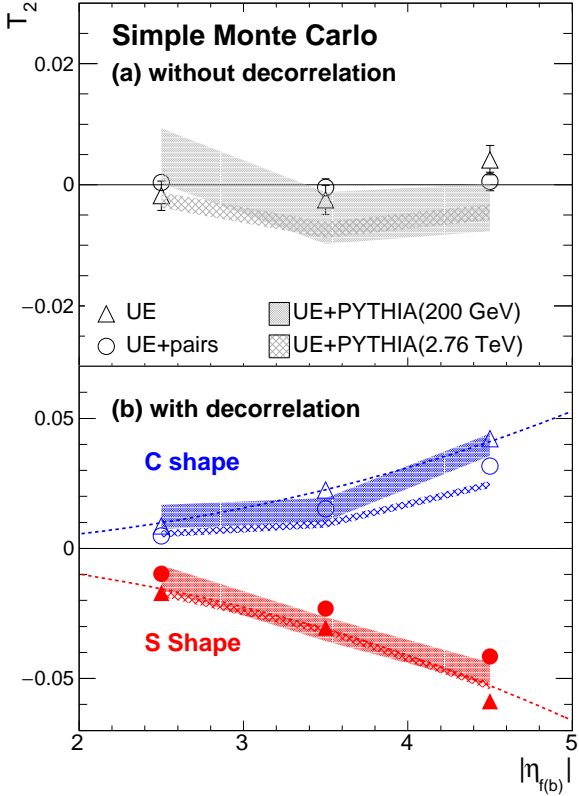}
\caption{\label{fig:T2_nonflow} (Color online) Simple Monte Carlo calculations of  $T_2$ as a function of $|\eta_{\rm f(b)}|$ {\color{black} for the underlying event (UE) without nonflow as well as for different scenarios of nonflow}.
$\frac{d\Delta\Psi}{d\eta}$ is set to $2^\circ$  for the S-shaped and the C-shaped flow-plane decorrelations.
The curves manifest Eqs.~\ref{eq:Sshape_T2} and~\ref{eq:Cshape_T2}  without nonflow. {\color{black}The shaded bands represent the two cases of PYTHIA embedding.} }
\end{figure} 

Defined as a four-particle cumulant, $T_2$ is essentially free from non-flow contribution (see Ref.~\cite{Aihong} and references therein).
$T_2$, as expressed in Eq.~\ref{eq:T2}, provides an intuitive way to tell whether
$\Psi_{\rm f}$ and $\Psi_{\rm b}$ fall on the same side or the
opposite sides of $\Psi_{\rm m}$:
a positive $T_2$ means a bow or a C-shaped decorrelation, and
a negative $T_2$ signifies a torque or an S-shaped decorrelation.
As done in the past, we shall exploit the linear-$\Delta\Psi(\eta)$
simulation without nonflow to learn the qualitative dependence of $T_2$ on $|\eta_{\rm  f(b)}|$ and $\frac{d\Delta\Psi}{d\eta}$ in different
decorrelation patterns. With a specific $\frac{d\Delta\Psi}{d\eta}$,
Fig.~\ref{fig:T2_MC} shows a rapid decreasing trend of $T_2$ vs
$|\eta_{\rm f(b)}|$ for the S-shaped case in panel (a), and an
increasing trend for the C-shaped case in panel (b). The simulated
points obey the following mathematical relations:
\begin{eqnarray}
T_2 &=& -\frac{\sin^2(2\frac{d\Delta\Psi}{d\eta} |\eta_{\rm f(b)}|)+\frac{1}{2}\cos(4\frac{d\Delta\Psi}{d\eta} |\eta_{\rm f(b)}|)}{\cos^2(2\frac{d\Delta\Psi}{d\eta} |\eta_{\rm f(b)}|)} + \frac{1}{2} \nonumber \\
&=& -\frac{1}{2}\tan^2(2\frac{d\Delta\Psi}{d\eta} |\eta_{\rm f(b)}|)  {\rm~(S~shape)} , \label{eq:Sshape_T2}\\
T_2 &=& \frac{\sin^2[2\frac{d\Delta\Psi}{d\eta}( |\eta_{\rm f(b)}|-0.5)]-\frac{1}{2}}{\cos^2[2\frac{d\Delta\Psi}{d\eta}( |\eta_{\rm f(b)}|-0.5)]}+\frac{1}{2} \nonumber \\
&=&\frac{1}{2}\tan^2[2\frac{d\Delta\Psi}{d\eta}( |\eta_{\rm f(b)}|-0.5)]
{\rm~(C~shape)} \label{eq:Cshape_T2}.
\end{eqnarray}
Again, in reality, the tilt angle may not increase linearly with the
$\eta$ gap, but we have confirmed with various monotonic function forms that the falling and rising trends of $T_2$ vs
$|\eta_{\rm f(b)}|$ should be solid expectations for the S-shaped and
the C-shaped decorrelations, respectively.

In $v_2\{\Psi_{\rm f(b)}\}$ and $F_2$ analyses, the core element is a cosine
function that yields large values close to 1 in strong-nonflow
scenarios. Conversely, $T_2$ uses the cumulant of a sine function that
gives close-to-zero nonflow contributions.  
Nonflow studies on $T_2$ are presented in Fig.~\ref{fig:T2_nonflow} with the same procedure as before. Panel (a) shows that for the scenario without any decorrelation, {\color{black}the $T_2$ results are mostly consistent with zero, with a potential of slightly negative values with the embedding of PYTHIA events at 2.76 TeV.} 
Panel (b) shows that for the scenarios with the C-shaped and the S-shaped decorrelations, the $T_2$ magnitude could be slightly reduced by nonflow, but the original trends are not changed vs $|\eta_{\rm f(b)}|$.

\section{AMPT studies}

\begin{figure}[tb]
\includegraphics[width=0.48\textwidth]{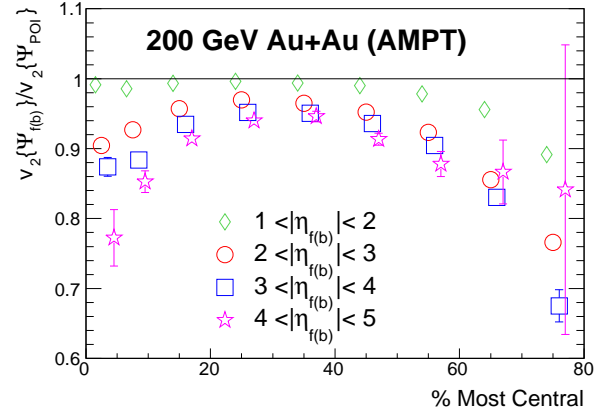}
\caption{\label{fig:v2_comp} (Color online) AMPT calculations of the $|\eta_{\rm f(b)}|$ dependence of the $v_2\{\Psi_{\rm f(b)}\}/v_{2}\{\Psi_{\rm POI}\}$ ratio in different centrality
intervals of Au+Au collisions at $\sqrt{s_{\rm NN}} =
  200$ GeV. Results using $\Psi_{\rm f}$ and $\Psi_{\rm b}$ are combined.  Some points are staggered horizontally to
  improve visibility.}
\end{figure}

We test the aforementioned methodology with more realistic events, simulated by the  AMPT model~\cite{ampt_1}.  AMPT is a hybrid transport event
generator, and describes four major stages of a high-energy heavy-ion
collision: the initial conditions, the partonic evolution, the
hadronization, and the hadronic interactions. For the initial
conditions, AMPT uses the spatial and momentum distributions of
minijet partons and excited soft strings, as adopted in the Heavy Ion
Jet Interaction Generator (HIJING)~\cite{ampt_2}.  Then Zhang's parton
cascade~\cite{ampt_3} is deployed to manage the partonic
evolution, determined by the two-body parton-parton elastic scattering.
At the end of the partonic evolution, the hadronization is implemented
via the spatial quark coalescence. 
Finally, the hadronic interactions are
modeled by a relativistic transport calculation~\cite{ampt_4}.
The string-melting (SM) version of AMPT reasonably well reproduces particle spectra and
elliptic flow in Au+Au collisions at $\sqrt{s_{\rm NN}} = 200$ GeV and Pb+Pb collisions at
2.76 TeV~\cite{ampt_5}.  In this study, Au+Au collisions at
 200 GeV are simulated by the SM 
version v2.25t4cu of AMPT.

\begin{figure}[tb]
\includegraphics[width=0.48\textwidth]{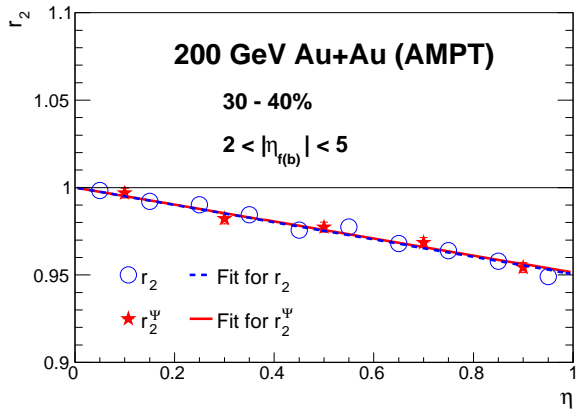}
\caption{\label{fig:r2_fig} (Color online) $r_2(\eta)$ and
  $r_2^{\Psi}(\eta)$ in 30-40\% Au+Au collisions at $\sqrt{s_{\rm NN}} =
  200$ GeV, simulated by AMPT. Results using $\Psi_{\rm f}$ and
  $\Psi_{\rm b}$ are combined. The straight-line fits to the $r_2$ and
  $r_2^\Psi$ points extract the corresponding $F_{2}$ of $(2.47 \pm
  0.11) \times 10^{-2}$ and $F_2^\Psi$ of ($2.43 \pm 0.11) \times
  10^{-2}$.  }
\end{figure}

\begin{figure}[b]
\includegraphics[width=0.48\textwidth]{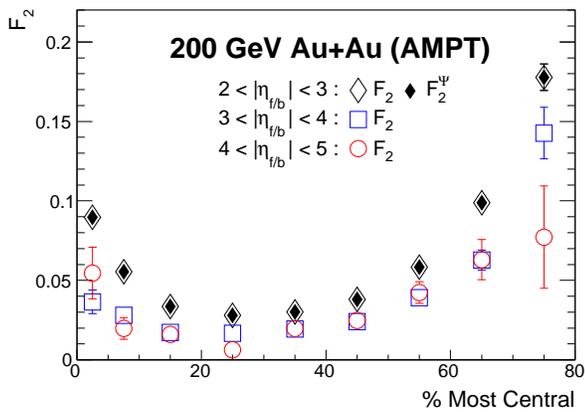}
\caption{\label{fig:F2_fig} (Color online) $F_2$ and $F_2^{\Psi}$ {\color{black} for different $\eta_{\rm f/b}$ ranges} as  functions of centrality in Au+Au collisions at $\sqrt{s_{\rm NN}} =
  200$ GeV from AMPT.  }
\end{figure}

In the following analyses of the AMPT events, we only select $\pi^\pm$,
$K^\pm$, $p$ and $\bar p$ with $0.15 < p_T < 2$ GeV/$c$.
$\varphi_{\rm m}$ and $\Psi_{\rm m}$ are delimited with $|\eta|<1$,
and $\Psi_{\rm f}$ and $\Psi_{\rm b}$ are reconstructed with particles
within $2<\eta<5$ and $-5<\eta<-2$, respectively.
With the expectations from nonflow and the scenarios of flow-plane
decorrelations in mind (see Figs.~\ref{fig:v2_MC} and~\ref{fig:v2_nonflow}), we examine the $|\eta_{\rm f(b)}|$
dependence of the $v_2\{\Psi_{\rm f(b)}\}/v_2\{\Psi_{\rm POI}\}$ ratio in AMPT.
Figure~\ref{fig:v2_comp} shows the AMPT calculations of
this ratio for four $|\eta_{\rm f(b)}|$ bins in
different centrality intervals of Au+Au collisions at $\sqrt{s_{\rm NN}} =
  200$ GeV. Within each centrality bin, $v_2\{\Psi_{\rm f(b)}\}/v_2\{\Psi_{\rm POI}\}$ displays
a decreasing trend with the increasing $\eta$ gap, excluding the S-shaped decorrelation from the dominant underlying
mechanisms.  Both nonflow and the C-shaped decorrelation could induce
such a decreasing trend in the $v_2$ ratio, and we cannot yet discern the two scenarios using this observable. 

Next, the AMPT results of $r_2(\eta)$ and
$r_2^{\Psi}(\eta)$ are compared in Fig.~\ref{fig:r2_fig} for 30-40\% Au+Au collisions at 200 GeV.  As mentioned earlier, $\Psi_{\rm f}$ can be replaced with
$\Psi_{\rm b}$ in the $r_2$ definition, with $\eta$ and $-\eta$ simultaneously exchanged. The two sets of results have been combined to
gain better statistics.
Both $r_2$ and $r_2^\Psi$ show a  linear trend decreasing as
$\eta$ increases, seemingly indicating a torque in the flow-plane
decorrelation at midrapidities.  The linear fits render the very close values of $F_{2} = (2.47 \pm 0.11) \times 10^{-2}$ and
$F_2^\Psi = (2.43 \pm 0.11)\times 10^{-2}$, which implies a marginal contribution of $v_2$-magnitude decorrelations  at midrapidities in AMPT events.

\begin{figure}[tb]
\includegraphics[width=0.48\textwidth]{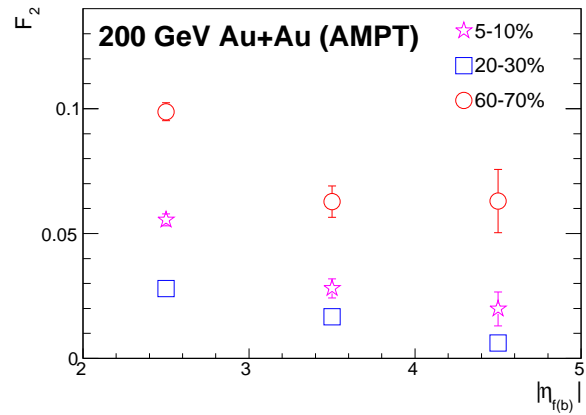}
\caption{\label{fig:F2_eta_pt} (Color online) AMPT calculations of
  $F_2$ versus $|\eta_{\rm f(b)}|$ for three selected centrality ranges in Au+Au collisions at $\sqrt{s_{\rm NN}} =
  200$ GeV.}
\end{figure}

Figure~\ref{fig:F2_fig} presents the AMPT calculations of $F_2$ and $F_2^\Psi$ {\color{black} for different $\eta_{\rm f/b}$ ranges} as a function of centrality. The two quantities are very close to each other in most centrality bins, {\color{black} and to avoid clutter, we only show $F_2^\Psi$ for the case of $2<|\eta_{\rm f/b}|<3$ as a demonstration}.
Thereafter, we  focus on $F_2$.  {\color{black} For the three $|\eta_{\rm f/b}|$ ranges,} the magnitude of $F_2$ is about 2.5\% in the 10--50\% centrality range,
and becomes larger in more central and more peripheral events.  Before
attributing the finite $F_2$ values to the flow-plane decorrelation,
one should note that such a centrality dependence could be a
reflection of nonflow effects. Nonflow contributions are positive in
$F_2$, and become more pronounced in peripheral collisions where
multiplicity is low and in central collisions where $v_2$ is small. This
caveat also applies to experimental data~\cite{Maowu, rn} that show $F_2$ features similar to these AMPT
simulations.
{\color{black}Moreover, the $F_2$ dependence on $|\eta_{\rm f/b}|$ and centrality qualitatively resembles Fig.2(a) of Ref.~\cite{ATLAS2}, corroborating the nonflow contribution.
}

The qualitative expectation from the simple Monte Carlo simulation (see Figs.~\ref{fig:F2_MC} and~\ref{fig:F2_nonflow})
motivates the differential measurements of $F_2$ with respect to
$|\eta_{\rm f(b)}|$. Figure~\ref{fig:F2_eta_pt}
delineates AMPT calculations of $F_2$ as a function of $|\eta_{\rm f(b)}|$ for three selected centrality bins in Au+Au collisions at 200 GeV. The $F_2$ values
are positive for all the cases under study, consistent with the
S-shaped decorrelation. However, the $|\eta_{\rm f(b)}|$ dependence
shows a falling trend  in each centrality
interval, in contrast to the rising trend expected for the torque alone.  Therefore, the contribution of the S-shaped flow-plane decorrelation in these $F_2$ values, if any, must have been
dominated by nonflow effects, which are also positive, and decrease with increasing $|\eta_{\rm f(b)}|$.

\begin{figure}
\includegraphics[width=0.48\textwidth]{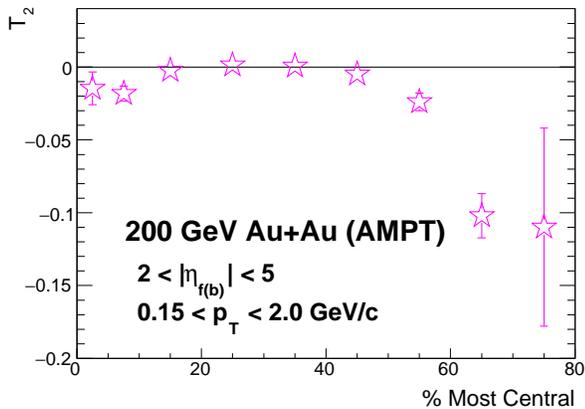}
\caption{\label{fig:t2_full} (Color online) Centrality dependence of $T_2$ in AMPT events of Au+Au collisions at $\sqrt{s_{\rm NN}} =
  200$ GeV. The $|\eta_{\rm f(b)}|$ range is (2, 5) for $\Psi_{\rm f}$ and $\Psi_{\rm b}$. 
}
\end{figure}

\begin{figure}
\includegraphics[width=0.48\textwidth]{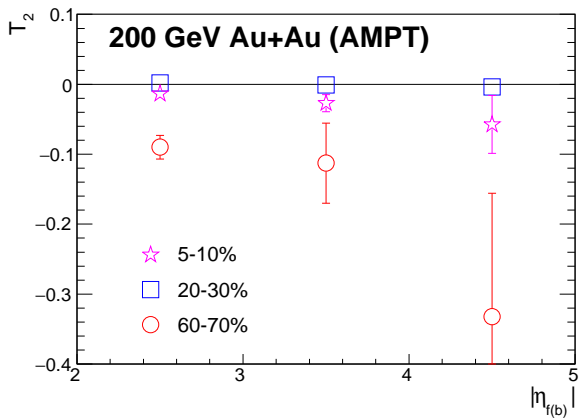}
\caption{\label{fig:T2_eta_pt} (Color online) $T_2$ dependence on $|\eta_{\rm f(b)}|$ for three selected centrality ranges in AMPT events of Au+Au at $\sqrt{s_{\rm NN}} =
  200$ GeV.}
\end{figure}

Finally, Fig.~\ref{fig:t2_full} depicts the $T_2$ values as a function of
centrality in the AMPT events of Au+Au collisions at
200 GeV. For the 10--50\% centrality range, $T_2$ is consistent with zero, indicating no apparent flow-plane decorrelation. In more central or more peripheral events, $T_2$ falls negative, supporting the picture of a torque or the S-shaped decorrelation.
Note that when the
tilt angle between $\Psi_{\rm m}$ and $\Psi_{\rm f(b)}$ becomes large,
the denominator of $T_2$ significantly under-estimates the product of
the event-plane resolutions, which in turn will blow up the magnitudes of both $T_2$ and
its statistical uncertainty.  This effect partially attributes to the large $T_2$ magnitudes and
the large error bars in peripheral collisions. The bright side is that
this effect does not change the sign of $T_2$, because the denominator is always positive by symmetry.

The AMPT results of $T_2$ as a function of $|\eta_{\rm f(b)}|$ are
plotted in Fig.~\ref{fig:T2_eta_pt} for three selected centrality bins
in Au+Au collisions at 200 GeV. For the 20--30\% centrality range, $T_2$ is always consistent with zero, whereas for 5--10\% and 60--70\% collisions, there seem to be decreasing trends due to a torque, although the statistical uncertainties become large at increased $|\eta_{\rm f(b)}|$.  Therefore, both the sign and the $|\eta_{\rm
  f(b)}|$-dependence of $T_2$ confirm the S-shaped longitudinal
decorrelation in central and peripheral AMPT events. 

A comparison between $F_2$ and $T_2$ is 
possible, if we neglect nonflow effects for the moment, and use Eqs.~\ref{eq:F2_eta},~\ref{eq:Sshape_T2} and \ref{eq:Cshape_T2} to extract the tilt angles from these two observables. Table~I lists the values of $F_2$ and $T_2$ as a function of centrality {\color{black} for $2<|\eta_{\rm f(b)}|<5$} from Figs.~\ref{fig:F2_fig} and~\ref{fig:t2_full}, as well as the extracted tilt angles. The average $|\eta_{\rm f(b)}|$ is around 3.2 with a weak centrality dependence. In general, the tilt angles obtained from $F_2$ have  magnitudes larger than those from $T_2$, and the difference could be attributed to  contributions other than nonflow.
The smaller  tilt-angle values obtained from $T_2$ might also indicate a stochastic nature of the flow-plane decorrelations as a function of $\Delta\eta$. For example, one can model the flow-plane angle along the longitudinal direction with a Markov chain: the $\eta$ range is divided into many small steps, and the flow plane at each step is randomly tilted by a small amount with respect to that at the previous step. In that case $\mean{(\Delta\Psi)^2} \propto \Delta\eta$.  
Such a random-walk-like process will lead to a positive $F_2$ independent of $|\eta_{\rm f(b)}|$, but have zero contribution to $T_2$.

\begin{table}
\caption{AMPT results of $T_2$ and $F_2$ {\color{black} for $2<|\eta_{\rm f(b)}|<5$} from Figs.~\ref{fig:F2_fig} and~\ref{fig:t2_full} as well as the corresponding tilt angles ($\frac{d\Delta\Psi}{d\eta}$) versus centrality in Au+Au collisions at 200 GeV. 
$\frac{d\Delta\Psi}{d\eta}$ is extracted using Eqs.~\ref{eq:F2_eta},~\ref{eq:Sshape_T2} and \ref{eq:Cshape_T2} without nonflow. For the tilt angles from $T_2$, a minus sign is added, if the decorrelation is C-shaped.}
\begin{tabular}{ c || c | c| c | c}
\hline
\% central & $F_2$ (\%) & $T_{2}$ (\%) & $\frac{d\Delta\Psi}{d\eta}\{F_2\}$ ($^\circ$) &$\frac{d\Delta\Psi}{d\eta}\{T_2\}$ ($^\circ$) \\
\hline
0-5 & 7.0 (0.4) & -1.5 (1.1) & 4.26 (0.13) & 1.3 (0.9)
\\
\hline
5-10 & 4.3 (0.2) & -1.8 (0.5) & 3.37 (0.09) & 1.7 (0.2) 
\\
\hline
10-20 & 2.6 (0.1) & -0.2 (0.2) & 2.61 (0.06) & 0.5 (0.4) 
\\
\hline
20-30 & 2.1 (0.1) & 0.1 (0.2) & 2.35 (0.06) & -0.4 (0.5) 
\\
\hline
30-40 & 2.6 (0.1) & 0.1 (0.2) & 2.57 (0.06) & -0.2 (0.6) 
\\
\hline
40-50 & 3.1 (0.1) & -0.5 (0.3) & 2.84 (0.06) & 0.8 (0.3)
\\
\hline
50-60 & 4.9 (0.2) & -2.4 (0.6) & 3.57 (0.08) & 1.9 (0.2) 
\\
\hline
60-70 & 8.1 (0.4) & -10.2 (1.5) & 4.58 (0.10) & 3.8 (0.3) 
\\
\hline
70-80 & 16.0 (0.9) & -11.0 (6.8) & 6.42 (0.17) & 3.6 (1.7) 
\\
\hline
\end{tabular}
\label{tab:Ntrans}
\end{table}

\section{Conclusions}

The flow-plane azimuthal decorrelations provide important input to the
initial condition and the system evolution of heavy-ion collisions in
the longitudinal dimension. We have explored three analyses that may be sensitive to such decorrelations:
$v_2\{\Psi_{\rm f(b)}\}/v_2\{\Psi_{\rm POI}\}$, $F_2^{(\Psi)}$  and $T_2$. 
Using simple Monte Carlo simulations with a constant $\frac{d\Delta\Psi}{d\eta}$ slope, we have learned the impacts of decorrelation and nonflow on these observables as functions of $|\eta_{\rm f(b)}|$.
\begin{itemize}
\item In the absence of nonflow, the $v_2\{\Psi_{\rm f(b)}\}/v_2\{\Psi_{\rm POI}\}$ ratio will increase(decrease) with increasing $|\eta_{\rm f(b)}|$, if the decorrelation is S-shaped(C-shaped). However, nonflow tends to pull down the trend, and could cause a falling trend even with the S-shaped decorrelation.

\item Without nonflow, $F_2$ is zero in the case of the C-shaped decorrelation or no decorrelations, and is positive, increasing with $|\eta_{\rm f(b)}|$ in the S-shaped case. Again, nonflow could force a decreasing trend regardless of whether the decorrelation is a torque or a bow.

\item $T_2(|\eta_{\rm f(b)}|)$ is expected to distinguish 
the C-shaped and the S-shaped decorrelaton patterns with opposite trends.
Nonflow could slightly reduce the magnitude of $T_2$, but is unlikely to  change the pertinent trend.

\end{itemize}

We have further applied the aforementioned methods to the AMPT events of Au+Au collisions at $\sqrt{s_{\rm NN}} = 200$ GeV. The findings are  summarized in the following.
\begin{itemize}
\item For each centrality interval under study, the $v_2\{\Psi_{\rm f(b)}\}/v_2\{\Psi_{\rm POI}\}$ ratio decreases with increasing $|\eta_{\rm f(b)}|$, which means that the S-shaped decorrelation cannot be the dominant underlying mechanism for this observable. This approach is unable to separate nonflow from the C-shaped decorrelation.

\item $F_2$ and $F_2^{\Psi}$ are very close to each other in most cases, indicating a negligible $v_2$-magnitude decorrelation in
AMPT. It is of interest to see if that is also true in experimental  data.

\item Although the positive sign of $F_2^{(\Psi)}$ seems to evidence the torque in AMPT, nonflow contributions
cast a shadow over this interpretation, with the decreasing
trend of $F_2(|\eta_{\rm f(b)}|)$. 

\item The novel observable, $T_2$, suppresses nonflow with the sine
function and the cumulant treatment. $T_2$ is consistent with no decorrelation for 10--50\% AMPT events of Au+Au at 200 GeV, and supports the S-shaped decorrelation picture in
0--10\% and 50--80\% collisions, with both the {\color{black} right} sign and the {\color{black} expected} $|\eta_{\rm f(b)}|$-dependence.
\end{itemize}

We have demonstrated the importance of the $|\eta_{\rm f(b)}|$ dependence of the experimental observables for the interpretation in terms of flow-plane decorrelations. The proposed $T_2$ observable exhibits advantages over the $v_2$ ratio and $F_2$ with regard to nonflow  effects, and is  sensitive to  details of the decorrelation patterns tracing back to the initial participant matter. Although we concentrate on the $2^{\rm nd}$-harmonic flow, the methodology presented in this paper can be readily extended to higher harmonics. We look forward to the corresponding applications  to the 
real-data analyses.

\begin{acknowledgments}
The authors thank Zi-Wei Lin and Guo-Liang Ma for providing the AMPT code.
We thank Maowu Nie, Jiangyong Jia, Aihong Tang, ShinIchi Esumi and Mike Lisa for the
fruitful discussions.   Z. Xu, X. Wu, G. Wang and H. Huang are  supported
by the U.S. Department of Energy under Grant No. DE-FG02-88ER40424. C. Sword and S. Voloshin are supported by the U.S. Department of Energy Office of Science, Office of Nuclear Physics under Grant No. DE-FG02-92ER40713.
\end{acknowledgments}

\appendix
\section{Two-particle correlations in the simple simulation}
\label{Append:Nonflow}

\begin{figure}[tb]
\includegraphics[width=0.48\textwidth]{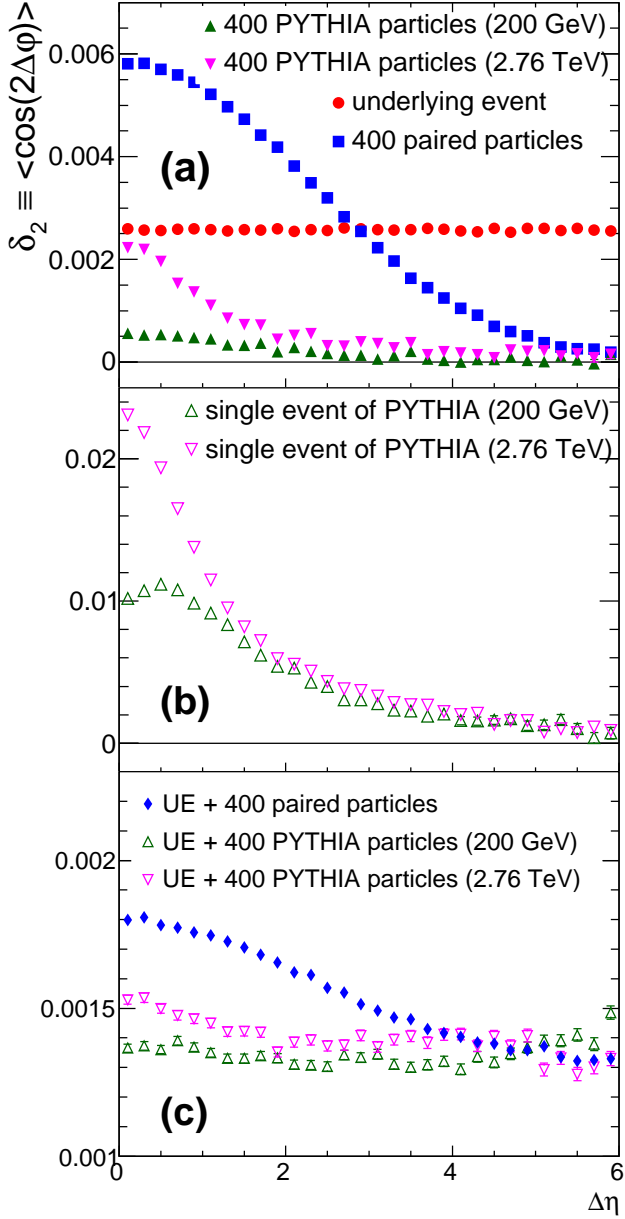}
\caption{\label{fig:appendix_15} (Color online) $\delta_2$ as a function of $\Delta \eta$ for different
classes of particles in the simple Monte-Carlo simulations.
Panel (a) includes the underlying event without nonflow, the simple nonflow implementation with 400 paired particles, and two cases of 400 PYTHIA particles from p+p collisions at $\sqrt{s} = 200$ GeV and 2.76 TeV, respectively.
The results from single PYTHIA events are presented in panel (b).}
\end{figure}

The nonflow effects can be easily visualized in the correlation function between two particles, $\delta_2 \equiv \langle \cos[2(\varphi_1 - \varphi_2)] \rangle$, as a function of their $\eta$ gap  ($\Delta\eta$).
Without nonflow or decorrelation, $\delta_2$ simply reflects $\langle v_2^2 \rangle$.
Figure~\ref{fig:appendix_15}(a) presents  $\delta_2(\Delta\eta)$ for four distinct classes of particles without decorrelation in the simple Monte Carlo simulation. The 1000 particles in each underlying event uniformly cover the $\eta$ range of $(-5,5)$, and carry an average $v_2$ of 5\%. From event to event, $v_2$ follows a uniform distribution from 3.5\% to 6.5\%. Since the underlying events are free of nonflow, the corresponding $\delta_2$ remains constant at $\langle v_2 \rangle^2+\sigma^2$, where $\sigma\sim0.75\%$ is due to the event-by-event fluctuation. The second class consists of 400 paired particles  without elliptic flow. Within each of the 200 pairs, the two particles have the same  azimuthal angle,
and their $\Delta\eta$ follows a Gaussian distribution with zero mean and a width of 2 units of $\eta$. Although the mean values of $\delta_2$ are almost identical for the 1000 underlying-event particles and the 400 paired particles, the $\Delta\eta$ dependence is very different between these two cases. The former is constant, whereas the latter is  enhanced at small $\Delta\eta$, and approaches zero at large $\eta$ gaps.
The result from the simple nonflow implementation qualitatively resembles the correlation between the 400 PYTHIA particles (triangular markers), with stronger effects. On average, the combination of 21(12) single PYTHIA events of p+p collisions at 200 GeV(2.76 TeV) provides the 400 PYTHIA particles. Although a single PYTHIA event has very strong nonflow effects as shown in panel (b), the correlation strength is diluted roughly by a factor of 21 or 12, when such a number of  PYTHIA events are merged together. 
Interestingly, the single PYTHIA events of different collision energies illustrate similar long-range correlations at $\Delta\eta>1$.
Finally, panel (c) shows the correlation results after the ``nonflow" particles are embedded into the underlying events. In general, the embedded events display lower $\delta_2$ values than the underlying events, since the 400 particles with no flow dilute  the 1000 flowing particles. Meanwhile,  nonflow correlations cause the nonuniform $\Delta\eta$ dependence. In practice, the simple Monte Carlo simulation can be tuned to compare with real data, and help us understand the nonflow contributions.

\section{Cumulant of four-angle correlations}
\label{Append:cumu}

The two-variable cumulant of an observable, $f = x y$, is defined as
\begin{equation}
\langle\langle xy \rangle\rangle \equiv \langle xy \rangle - \langle x \rangle \langle y \rangle.     
\end{equation}
The first term computes the average of the product, accounting for
correlations between $x$ and $y$, while the second gives the product
of the averages, keeping $x$ and $y$ independent of each other. A zero
cumulant means that $x$ and $y$ are independent with no correlation.
The $T_2$ observable involves four azimuthal angles, and therefore we
 derive the four-variable cumulant, which subtracts all
lower-order correlations:
\begin{eqnarray}
\langle\langle xyzt \rangle\rangle &\equiv&
\langle xyzt \rangle - \langle\langle xyz \rangle\rangle \langle t \rangle - \langle\langle xyt \rangle\rangle \langle z \rangle \nonumber\\
& & - \langle\langle xzt \rangle\rangle \langle y \rangle - \langle\langle yzt \rangle\rangle \langle x \rangle - \langle\langle xy \rangle\rangle \langle\langle zt \rangle\rangle 
\nonumber\\
& & -\langle\langle xz \rangle\rangle
 \langle\langle yt \rangle\rangle -  
 \langle\langle xt \rangle\rangle 
 \langle\langle yz \rangle\rangle \nonumber \\
 & & - \langle x \rangle\langle y \rangle\langle z \rangle\langle t \rangle.
 \label{4cumu}
\end{eqnarray}
In principle, we should obtain the three-variable cumulants before calculating the four-variable cumulant.
However, in the case of $T_2$, all the variables can be expressed as a sine or cosine function of an azimuthal angle, and we can force $\langle x \rangle=\langle y \rangle=\langle z \rangle=\langle t \rangle=0$ in the data analyses, e.g., via the shifting method~\cite{shifting}. Thus, Eq.~\ref{4cumu} is simplified to
\begin{equation}
\langle\langle xyzt \rangle\rangle =  \langle xyzt \rangle - \langle xy \rangle   \langle zt \rangle
- \langle xz \rangle   \langle yt \rangle
-\langle xt \rangle   \langle yz \rangle.
\label{eq:4cumu}
\end{equation}

The numerator of $T_2$ can be expanded into four terms:
\begin{eqnarray}
& &\sin(a-b)\sin(c-d) \nonumber\\
&=&  \sin(a) \cos(b) \sin(c) \cos(d) \nonumber\\
&&-\sin(a) \cos(b) \sin(d) \cos(c) \nonumber\\
&& - \sin(b) \cos(a) \sin(c) \cos(d) \nonumber\\
&& + \sin(b) \cos(a) \sin(d) \cos(c).
\label{eq:A4}
\end{eqnarray}
We put the first term into Eq.~\ref{eq:4cumu} as an example, 
\begin{eqnarray}
& &\langle\langle \sin(a) \cos(b) \sin(c) \cos(d)   \rangle\rangle  \nonumber \\
&=& \langle \sin(a) \cos(b) \sin(c) \cos(d)   \rangle \nonumber \\
& & - \langle \sin(a) \sin(c)   \rangle \langle \cos(b) \cos(d)   \rangle, \label{eq:A5}
\end{eqnarray}
where terms such as $\langle \sin(a) \cos(b) \rangle$ vanish in
symmetric heavy-ion collisions. Using the product-to-sum formulas, we
have
\begin{eqnarray}
& &\langle \sin(a) \sin(c)   \rangle \langle \cos(b) \cos(d)   \rangle \nonumber \\ 
&=& \frac{1}{4}[\langle \cos(a-c) \rangle \langle \cos(b-d) \rangle \nonumber \\
& &+ \langle \cos(a-c) \rangle \langle \cos(b+d) \rangle \nonumber \\
& & - \langle \cos(a+c) \rangle \langle \cos(b-d) \rangle \nonumber \\
& & \langle \cos(a+c) \rangle \langle \cos(b+d) \rangle].
\end{eqnarray}
Only the cosines with arguments of the angle difference are
independent of the coordinate system, and may render finite
averages. Hence Eq.~\ref{eq:A5} becomes
\begin{eqnarray}
& &\langle\langle \sin(a) \cos(b) \sin(c) \cos(d)   \rangle\rangle \nonumber \\ 
&=& \langle \sin(a) \cos(b) \sin(c) \cos(d)   \rangle \nonumber \\
& & - \frac{1}{4} \langle \cos(a-c) \rangle \langle \cos(b-d) \rangle. \label{eq:A7}
\end{eqnarray}
We follow the same procedure for the other three terms in
Eq.~\ref{eq:A4} and obtain
\begin{eqnarray}
& &\langle\langle
\sin(a-b)\sin(c-d)
\rangle\rangle \nonumber\\ &\equiv& \langle
\sin(a-b)\sin(c-d)
\rangle - \frac{1}{2}\langle
\cos(a-c)\rangle\langle\cos(b-d)
\rangle \nonumber\\
& &+ \frac{1}{2}\langle
\cos(a-d)\rangle\langle\cos(b-c)
\rangle.
\end{eqnarray}

\section{Generalization of \texorpdfstring{$T_2$}{Lg}}
\label{Append:general}

In the $T_2$ analyses in Sec.~\ref{Sec:T2}, we randomly split
particles within $|\eta|<1$ into two sub-events, and reconstruct
$\Psi_{\rm m,1}$ and $\Psi_{\rm m,2}$ based on them. By this means,
$\Psi_{\rm m,1}$ and $\Psi_{\rm m,2}$ are indistinguishable, sharing
the same kinematic region, bearing the same event plane resolution,
and tilting in the same way. In general, $\Psi_{\rm m,1}$ and
$\Psi_{\rm m,2}$ could come from different $\eta$ ranges, e.g., with
$-1<\eta_{\rm m,1}<0$ and $0<\eta_{\rm m,2}<1$. Accordingly, the
definition of $T_2$ is not unique any more, with a few possible
combinations.  For example, we can define
\begin{equation}
T_{2(I)} = \frac{\langle \langle \sin 2(\Psi_{\rm f} - \Psi_{{\rm m},2}) \sin 2(\Psi_{\rm b} - \Psi_{{\rm m},1})\rangle \rangle}{\langle \cos 2 (\Psi_{\rm f} - \Psi_{{\rm m},2}) \rangle \langle \cos 2(\Psi_{\rm b} - \Psi_{{\rm m},1})\rangle},
\end{equation}
and within the frame work of the simple Monte Carlo simulation, where
the tilt angle linearly increases with the $\eta$ gap, we have
\begin{eqnarray}
T_{2(I)} &=& -\frac{1}{2}\tan^2[2\frac{d\Delta\Psi}{d\eta}( |\eta_{\rm f(b)}|-0.5)]    {\rm~(S~shape)}, \label{eq:Sshape_T2I}\\
T_{2(I)} &=& \frac{1}{2}\tan^2[2\frac{d\Delta\Psi}{d\eta}( |\eta_{\rm f(b)}|-0.5)]
{\rm~(C~shape)} \label{eq:Cshape_T2I}.
\end{eqnarray}
We can switch $\Psi_{\rm m,1}$ and $\Psi_{\rm m,2}$ in $T_{2(I)}$ to
define a new observable
\begin{equation}
T_{2(II)} = \frac{\langle \langle \sin 2(\Psi_{\rm f} - \Psi_{{\rm m},1}) \sin 2(\Psi_{\rm b} - \Psi_{{\rm m},2})\rangle \rangle}{\langle \cos 2 (\Psi_{\rm f} - \Psi_{{\rm m},1}) \rangle \langle \cos 2(\Psi_{\rm b} - \Psi_{{\rm m},2})\rangle},
\end{equation}
and with the constant $\frac{d\Delta\Psi}{d\eta}$, we have
\begin{eqnarray}
T_{2(II)} &=& - \frac{1}{2}\tan^2[2\frac{d\Delta\Psi}{d\eta}( |\eta_{\rm f(b)}|+0.5)] {\rm~(S~shape)} , \label{eq:Sshape_T2II}\\
T_{2(II)} 
&=&\frac{1}{2}\tan^2[2\frac{d\Delta\Psi}{d\eta}( |\eta_{\rm f(b)}|-0.5)]
{\rm~(C~shape)} \label{eq:Cshape_T2II}.
\end{eqnarray}
A third type of $T_2$ observable can be defined as
\begin{equation}
T_{2(III)} = \frac{\langle \langle \sin 2(\Psi_{\rm f} - \Psi_{\rm b}) \sin 2(\Psi_{\rm m,1} - \Psi_{{\rm m},2})\rangle \rangle}{\langle \cos 2 (\Psi_{\rm f} - \Psi_{\rm b}) \rangle \langle \cos 2(\Psi_{\rm m,1} - \Psi_{{\rm m},2})\rangle},
\end{equation}
which is zero for the C-shaped decorrelation. With the constant $\frac{d\Delta\Psi}{d\eta}$, the expectation from the S-shaped case is 
\begin{equation}
T_{2(III)} = - \frac{1}{2}\tan(2\frac{d\Delta\Psi}{d\eta})\tan(4\frac{d\Delta\Psi}{d\eta} |\eta_{\rm f(b)}|). \label{eq:Sshape_T2III}    
\end{equation}

\end{document}